\documentclass[prl,twocolumn,showpacs,superscriptaddress,preprintnumbers,amssymb]{revtex4}
\usepackage{graphicx}
\usepackage{dcolumn}
\usepackage{bm}
\input epsf

\newcommand{\beq}{\begin{equation}}
\newcommand{\eeq}{\end{equation}}
\newcommand{\beqn}{\begin{eqnarray}}
\newcommand{\eeqn}{\end{eqnarray}}

\begin{document}
\title{Signatures of integrability in charge and thermal transport in 1D  quantum systems}
\author{Subroto Mukerjee}
\affiliation{Department of Physics, University of California,
Berkeley, CA 94720}
\affiliation{Materials Sciences Division,
Lawrence Berkeley National Laboratory, Berkeley, CA 94720}
\author{B. Sriram Shastry}
\affiliation{Department of Physics, University of California,
Santa Cruz, CA 95064}
\begin{abstract}
Integrable and non-integrable systems have very different
transport properties. In this work, we highlight these differences
for specific one dimensional models of interacting lattice
fermions using numerical exact diagonalization. We calculate the
finite temperature adiabatic stiffness (or Drude weight) and
isothermal stiffness (or ``Meissner'' stiffness) in electrical and
thermal transport and also compute the complete momentum and
frequency dependent dynamical conductivities $\sigma(q,\omega)$
and $\kappa(q,\omega)$. The Meissner stiffness goes to zero
rapidly with system size for both integrable and non-integrable
systems. The Drude weight shows signs of diffusion in the
non-integrable system and ballistic behavior in the integrable
system.  The dynamical conductivities are also consistent with
ballistic and diffusive behavior in the integrable and
non-integrable systems respectively.
\end{abstract}
\maketitle

A perfect metal is distinguished from a superconductor by the
absence of the Meissner effect, although both share an infinite
conductivity at zero frequency. In one dimension, another
fundamental kind of metallic systems is possible, corresponding to
an integrable system. We term this as an integrable metal (IM) and
such a system can display characteristics that are different from
both a perfect metal and a superconductor, such as an infinite
conductivity {\em at all temperatures}. In contrast, a perfect but
non integrable metal (NIM) becomes resistive at finite
temperatures due to inelastic collisions and umklapp. We track
this difference between an IM and NIM directly in this work by
calculating various transport stiffnesses also the dynamical
conductivities for both charge and energy using numerical exact
diagonalization and a suitably generalized Kubo
formula~\cite{shastry}. The main result is that the NIM shows
evidence of diffusive behavior in both the static and dynamic
conductivities, while the IM shows several signs of ballistic
transport.

The adiabatic stiffness (or Drude weight) is given
by~\cite{shastry}
\begin{equation}
\bar{D}_\alpha = \frac{1}{L}\left[\langle  \Gamma_\alpha \rangle -
\hbar \sum_{n ,m} (1-\delta_{\epsilon_n , \epsilon_m})
\frac{p_n-p_m}{\epsilon_m-\epsilon_n}|\langle n|J_\alpha|m
\rangle^2 \right], \label{stiff}
\end{equation}
Here $\Gamma_\alpha = - \lim_{k \rightarrow 0}
\frac{1}{k}[J_\alpha(k),K_\alpha(-k)]$, where $K_\alpha(k)$ is the
Fourier component of a local density operator ($\alpha=e$ for the
charge density, $\alpha=E$ for the energy density and $\alpha=Q$
for the heat density) while $J_\alpha(k)$ is that of the
corresponding current operator. Here $p_n = e^{-\beta
\epsilon_n}/Z$ is the Boltzmann weight of the many-body state with
energy $\epsilon_n$ and $Z$, the partition function. This is
arises in the adiabatic evolution of an equilibrium state
disturbed in the infinitely remote past by the application of a
time-dependent perturbation, and is the coefficient of the delta
function in the conductivity for charge transport.

Another stiffness, the isothermal (or ``Meissner'') stiffness
$D_\alpha$ can be defined by the same expression as
Eqn.~\ref{stiff} by omitting the Kronecker delta function that
forbids equal energy states. For charge transport, this arises in
a study of the Byers-Yang type curvature of the free energy with
respect to a flux through the ring~\cite{giamarchi}. On general
grounds, a superconductor which displays the Meissner effect has a
non-zero value of $D_e$. While a non-vanishing $D_\alpha$
implies a non-vanishing $\bar{D}_\alpha$, the converse is not
true.

The distinction between $D_\alpha$ and $\bar{D}_\alpha$ resides in
the relative statistical weight of states with equal energy. It
also follows that if $D_\alpha$ is known to vanish, then
$\bar{D}_\alpha$ can be written as a sum over only equal energy
states
\begin{eqnarray}\nonumber
\bar{D}_\alpha & = & \frac{\hbar}{k_B T L} \sum_{\epsilon_n =
\epsilon_m} p_n |\langle n|J_\alpha|m\rangle|^2 \\
&=&\bar{D}_\alpha^{d} + \bar{D}_\alpha^{nd} \label{stiffshort}
\end{eqnarray}
The current operator commutes with the crystal momentum, so $n$
and $m$ have the same value of crystal momentum $q$. In general,
an IM, can have degeneracies among states with same $q$ due to the
presence of dynamical symmetries. We can thus write $\bar{D}$ as
the sum of $\bar{D}_\alpha^{d}$, the contribution from states
degenerate with others {\it at the same value} of $q$ and
$\bar{D}_\alpha^{nd}$, the contribution from states with no
degeneracies {\it within the same} crystal momentum sector.

In an interesting paper~\cite{narozhny}, it has been argued that
in an IM with time reversal invariance, a non-zero value of
$\bar{D}_e$ is due to the presence of a large number of current
carrying states, which are degenerate with a partner state at the
opposite value of $q$. An important question is how these current
carrying states contribute to $\bar{D}_\alpha^{d}$ and
$\bar{D}_\alpha^{nd}$: What is the role of the dynamical
symmetries in determining the magnitude of the adiabatic
stiffness? This is especially relevant when time reversal is
broken by e.g. an irrational flux through the ring, as used here,
whereby even non-degenerate states {\it can carry currents}.

We have been motivated by these questions to undertake an
exhaustive numerical study of a typical IM and NIM, by computing
their exact energy spectra and all the current matrix elements and
consequently the transport stiffnesses and dynamic conductivities.
This exercise is done in 1D with a popular model, the $t-t'-V$
model of spinless fermions, where a single non-zero parameter $t'$
destroys the integrability, but lets the system remain a perfect
metal (an NIM). All stiffnesses are generally non-zero for finite
systems, and it is only the systematics of their size dependence
that gives us reliable information on the large size behavior. We
also study for the first time, the complete dynamic conductivity
$\sigma(q,\omega)$ and thermal conductivity $\kappa(q,\omega$) in
an IM and NIM, using an appropriate scheme for binning the
discrete data points to obtain continuous functions.

\begin{figure}[ht!]
$\begin{array}{c}
\epsfxsize = 3.0in \epsfysize = 2.0in \epsffile{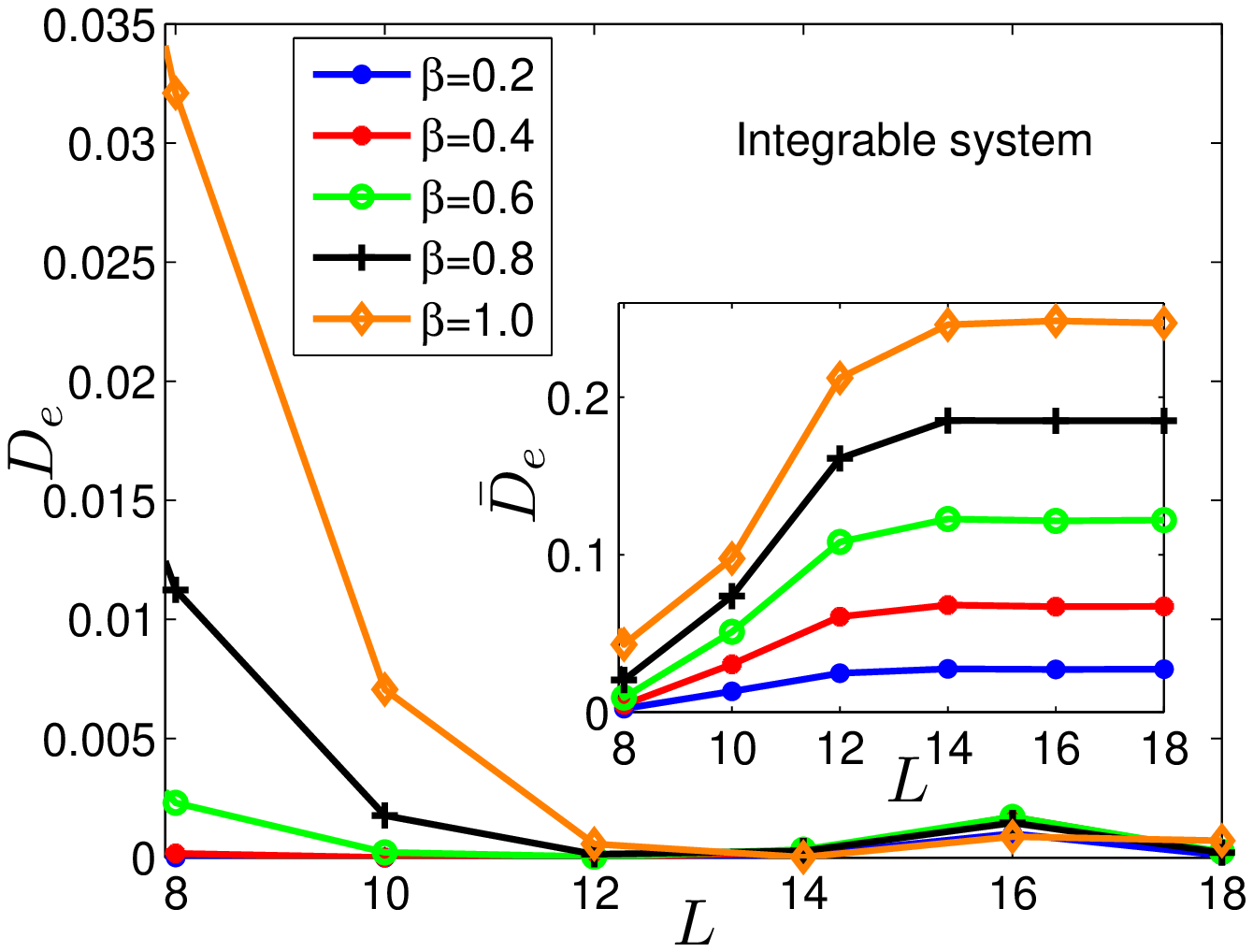} \\
\epsfxsize = 3.0in \epsfysize = 2.0in \epsffile{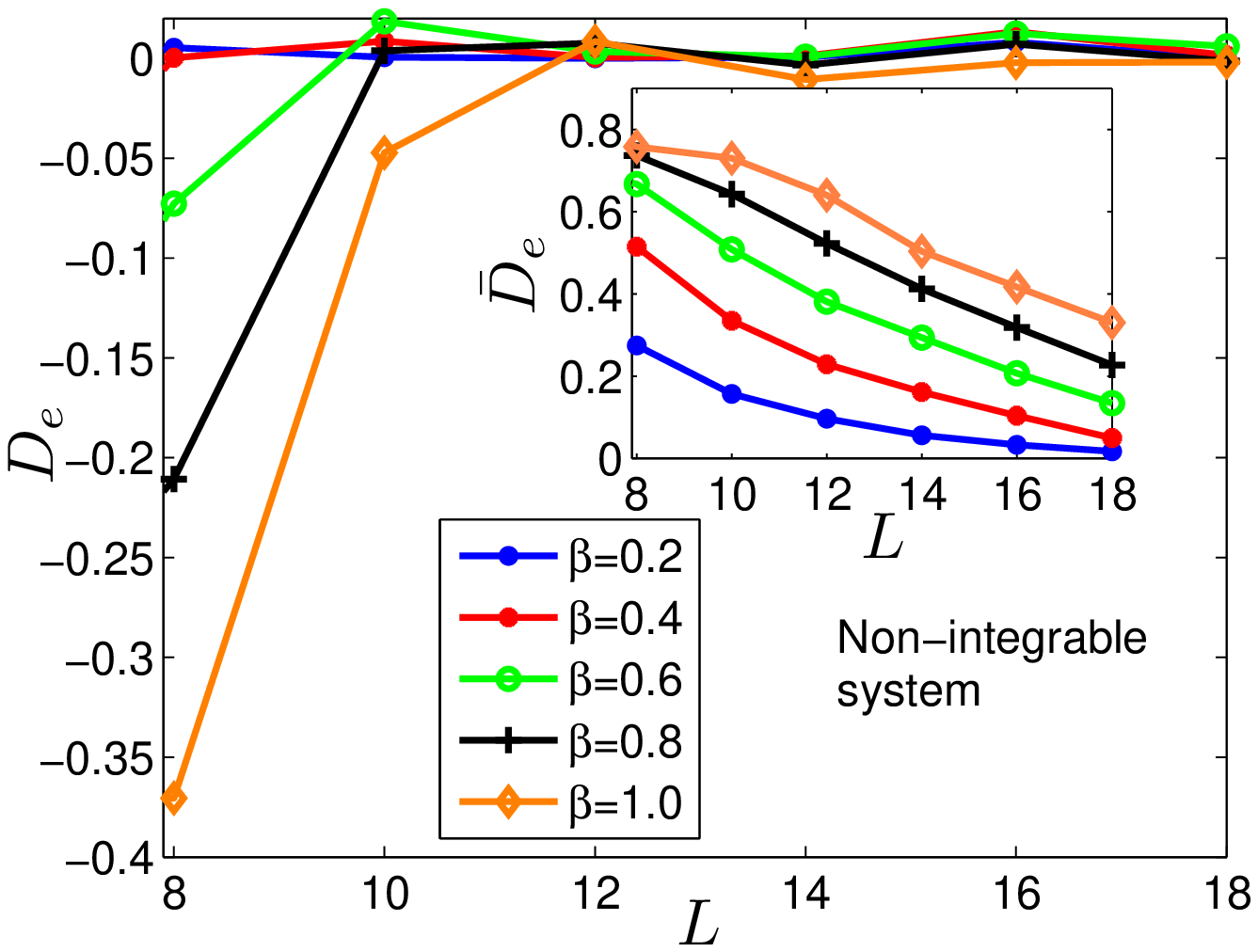}
\end{array}$
\caption{(Top) The isothermal and adiabatic (inset) stiffness for
electrical transport in the integrable system at various values of
$\beta$ with an irrational flux twist. It can be seen that the
former tends to zero rapidly with increasing system size while the
latter appears to go to a constant, indicative of ballistic
transport. (Bottom) The same stiffnesses for electrical transport
in the non-integrable system again with an irrational flux twist.
Both tend to zero with system size but $D_e$ does so more
rapidly.} \label{weights}
\end{figure}
\begin{figure}
$\begin{array}{c}
\epsfxsize = 3in \epsfysize = 2in \epsffile{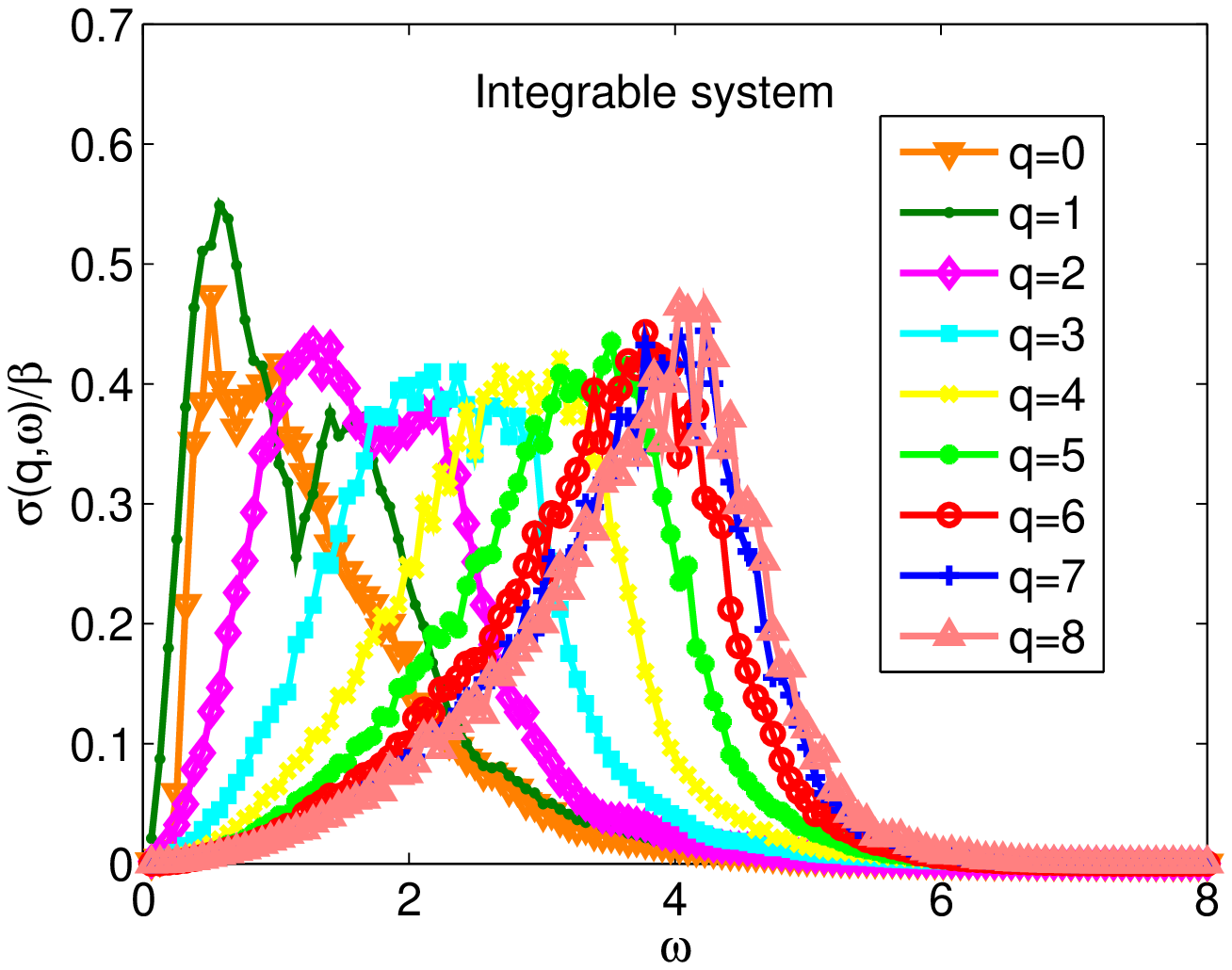} \\
\epsfxsize = 3in \epsfysize = 2in \epsffile{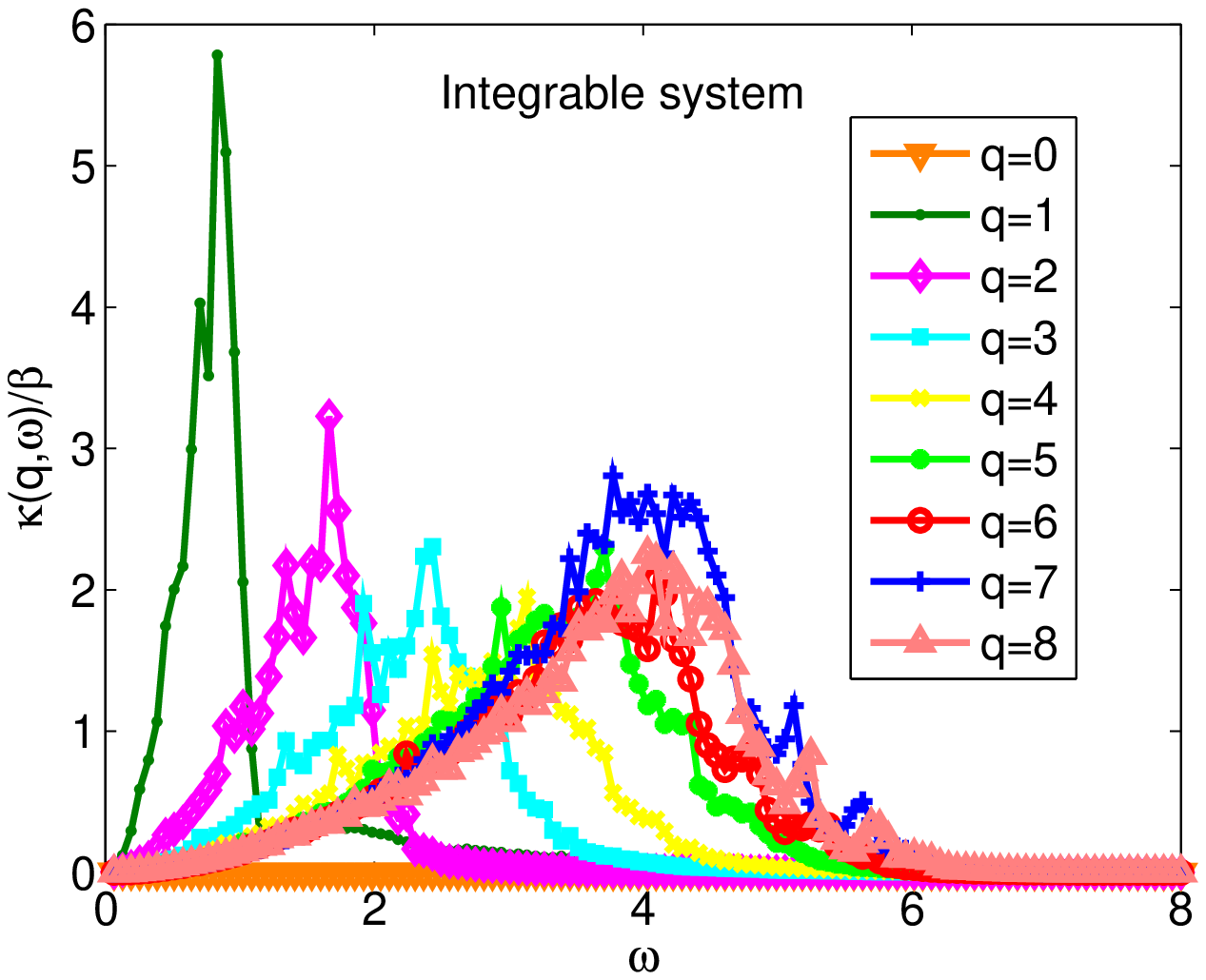}
\end{array}$
\caption{Plots of $\sigma(q, \omega)$ (above) and $\kappa(q,
\omega)$ (below) vs. $\omega$ for the integrable system with
$L=16$ at $\beta = 0.001$. It can be seen in the plot of
$\kappa(q,\omega)$ that there is a band of frequencies for a given
value of $q$ in which the value of $\kappa(q,\omega)$ is large and
then abruptly falls to zero at the boundaries. This behavior is
much less pronounced in $\sigma(q,\omega)$.} \label{diffmom}
\end{figure}
\begin{figure}
$\begin{array}{c}
\epsfxsize = 3in \epsfysize = 2in \epsffile{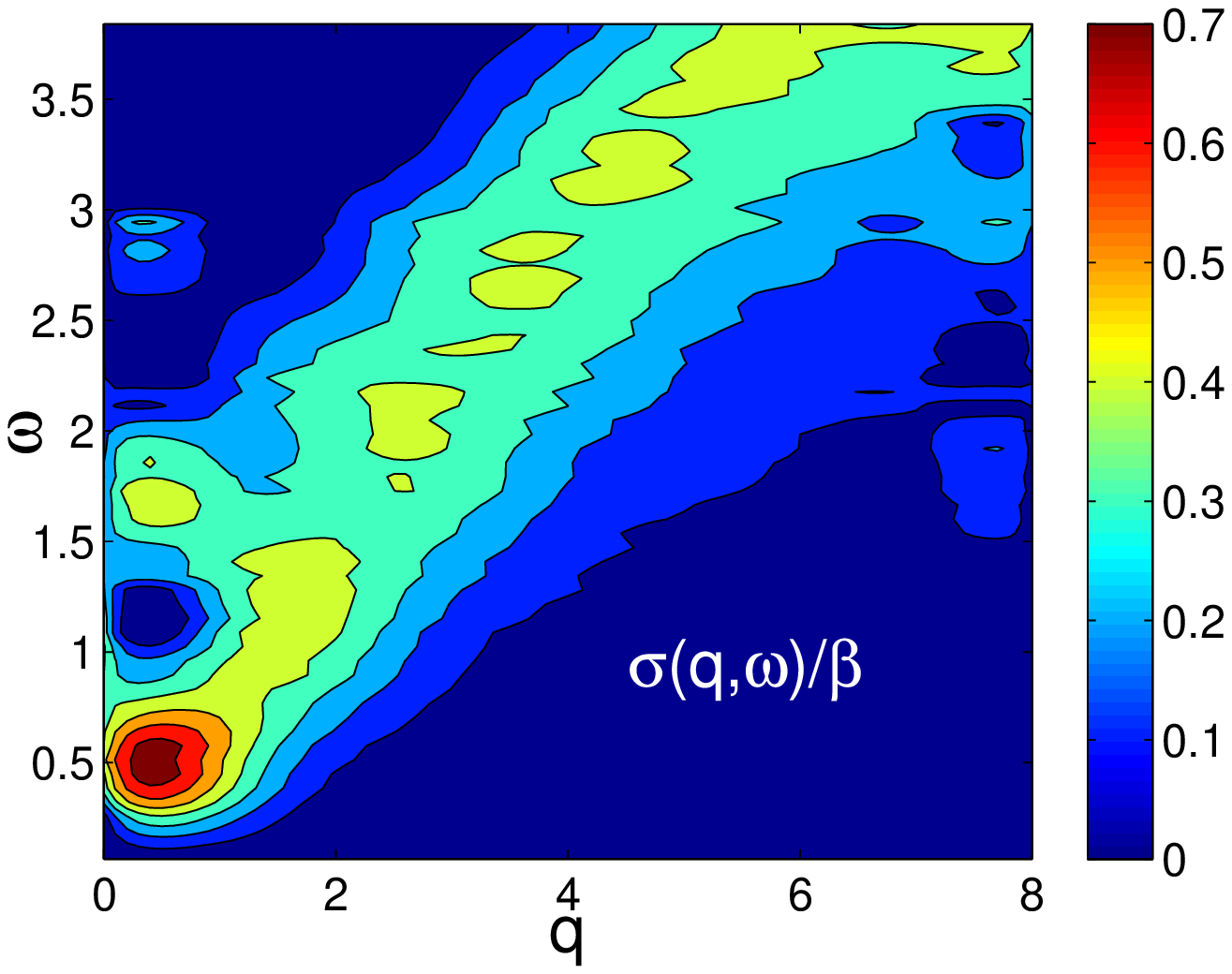} \\
\epsfxsize = 3in \epsfysize = 2in \epsffile{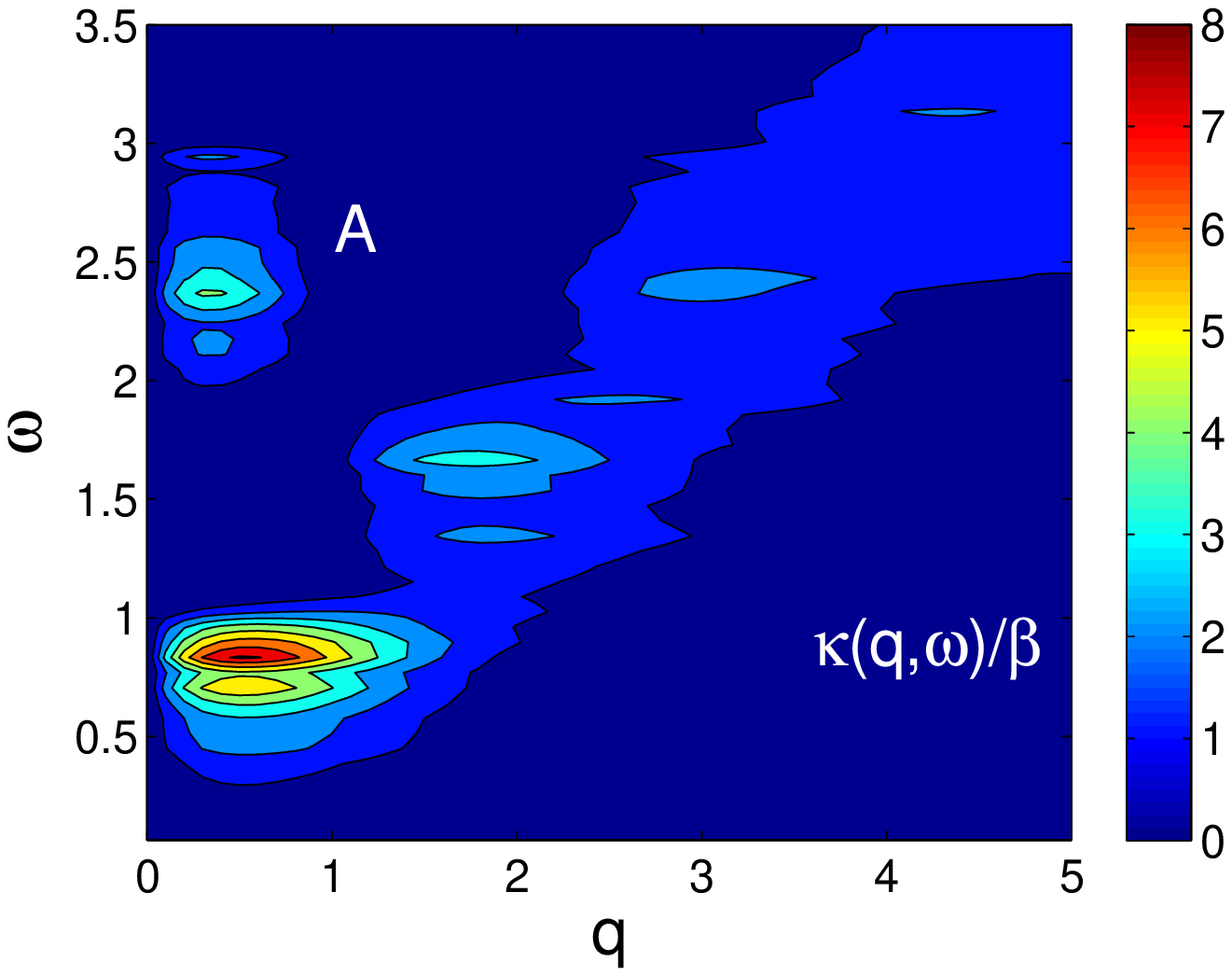}
\end{array}$
\caption{Colored contour plots of $\sigma(q, \omega)$ (above) and
$\kappa(q,\omega)$ (below) for the IM with $L=16$ at $\beta =
0.001$. Momentum ($q =$ 0 to 8). An eighth order polynomial
interpolation has been used to convert the discrete numerical data
into a function of a continuous momentum variable $q$ and the
feature marked A on the lower plot is an artifact of that. The
plot of $\kappa(q,\omega)$ shows the banded continuum in the
center bounded by dark blue regions.}\label{colcont}
\end{figure}
\begin{figure}
$\begin{array}{cc}
\epsfxsize = 3in \epsfysize = 2in \epsffile{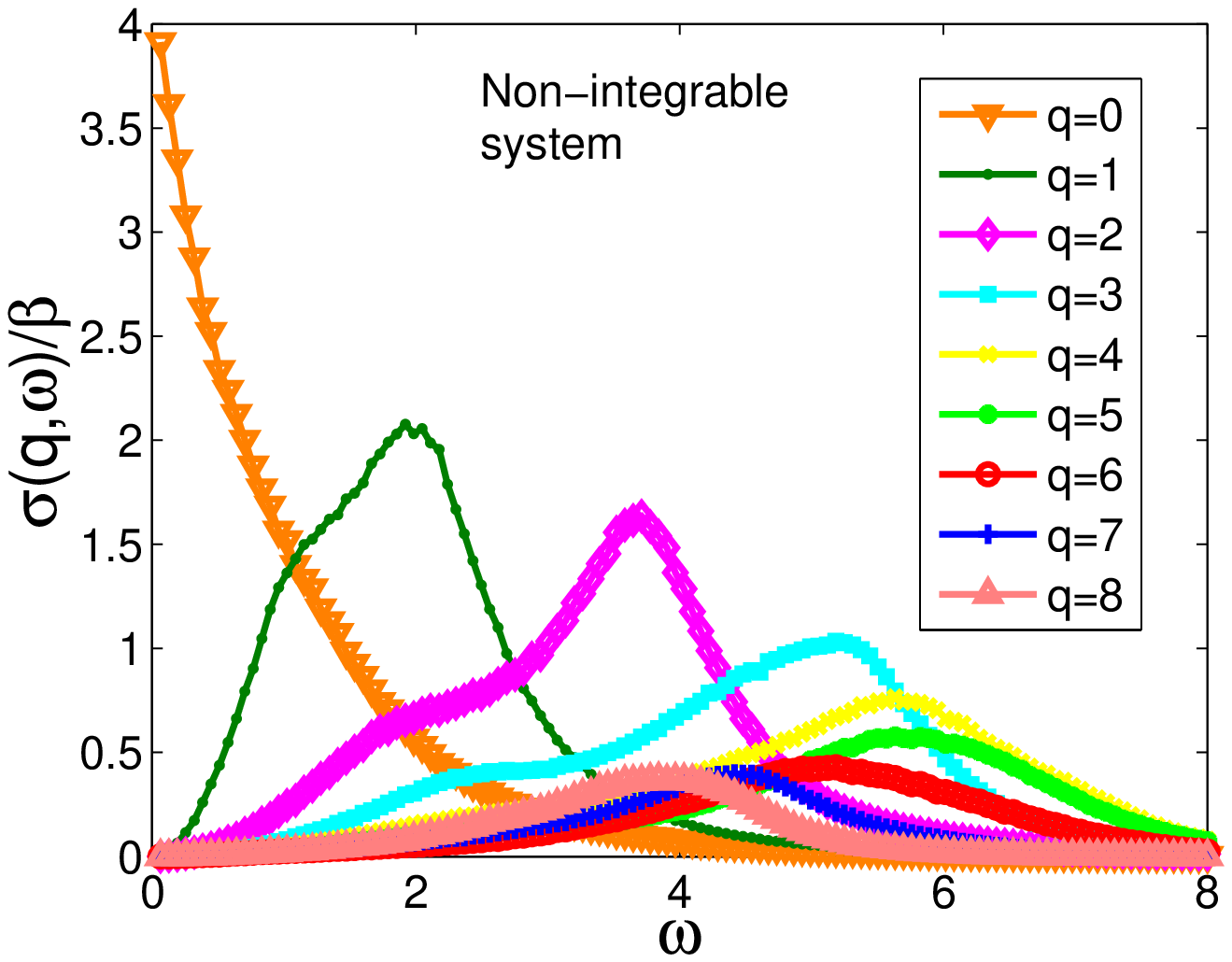} \\
\epsfxsize = 3in \epsfysize = 2in \epsffile{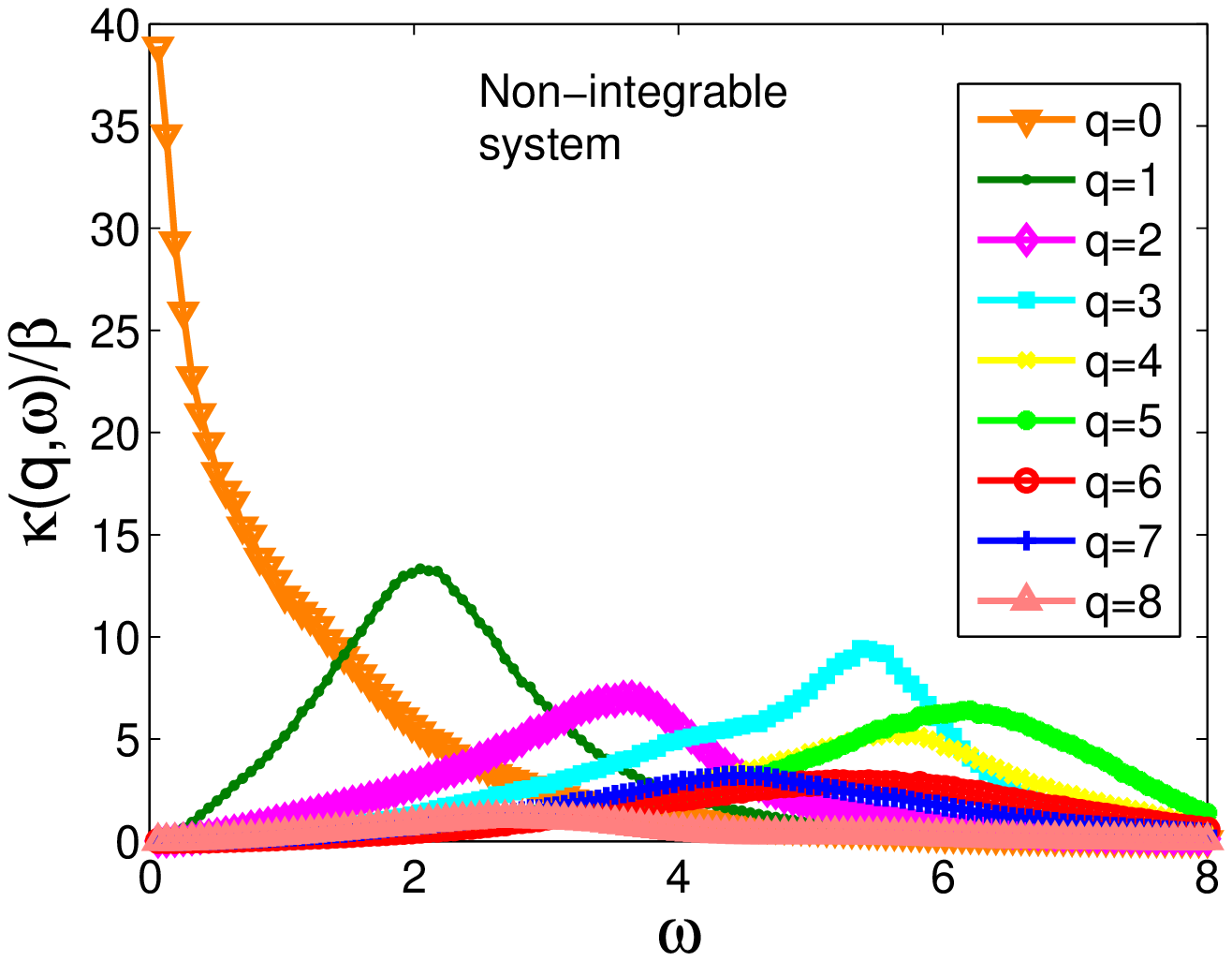}
\end{array}$
\caption{Plots of $\sigma(q,\omega)$ (above) and
$\kappa(q,\omega)$ (below) vs. $\omega$ for the NIM with $L=16$ at
$\beta = 0.001$. $\kappa(q,\omega)$ no longer displays the banded
behavior of the integrable system and falls off gradually (and not
abruptly) at higher frequencies. The peaks in both
$\sigma(q,\omega)$ and $\kappa(q,\omega)$ at large $q$ are also
present in the density-density correlation functions indicating
the presence of overdamped excitations.} \label{diffmomn}
\end{figure}

The $t-t'-V$ model of spinless fermions on a 1D ring is given by the Hamiltonian
\begin{eqnarray}
H & = & -t\sum_j(c_{j+1}^\dagger c_j + c_j^\dagger
c_{j+1})-t'\sum_j(c_{j+2}^\dagger c_j \nonumber \\
& & + c_j^\dagger c_{j+2}) + V\sum_j(n_j-1/2)(n_{j+1}-1/2).
\label{hamiltonian}
\end{eqnarray}
For concreteness, we choose the value of $t$ and $V$ to be 1.0 and
2.0 respectively. With $t'=0$, this model can be mapped onto the
integrable spin 1/2 Heisenberg ring using the Jordan-Wigner
transformation. We set $t'=1$ when we model the NIM. Chains of
length 6-18 are studied and we focus on half-filling for
definiteness. The Hamiltonian Eqn.~\ref{hamiltonian} is then
numerically diagonalized exactly. We would like to emphasize that
our conclusions vis a vis integrability and non-integrability are
quite general and independent of the microscopic parameters and
the filling.  We also apply an irrational flux twist to break
time-reversal invariance, which too does not affect the
integrability. {\em For convenience, we will henceforth use the
term non-degenerate only for states which are not degenerate with
any other state at the same $q$}. With the irrational flux twist,
the NIM now has only non-degenerate states while the IM has both
degenerate and non-degenerate states. This has the advantage of
enabling us to analyze the behavior of the IM in contrast to the
NIM, mainly in terms of degeneracies (and set $\bar{D}_e^{d}=0$
for the NIM). We calculate $D_\alpha$ and $\bar{D}_\alpha$ for
electrical and thermal transport at different temperatures
($T=1/\beta$ is measured in units of $t$).

We show here only the results for electrical transport in
Fig.~\ref{weights}, but the ones for thermal transport display
exactly the same qualitative behavior. The issue of whether the
charge stiffness ($\bar{D}_{e}$ in our model) in the Heisenberg
ring ($t'=0$) is actually non-zero at finite temperature remains
contentious~\cite{heisenberg,zotos2}. Our results up to the
largest system size ($L=18$) seem to indicate that it is non-zero.
Indeed, the overall picture that emerges is consistent with
ballistic transport. A confirmation of this to rule out any slow
(logarithmic) decay with $L$ is currently beyond the range of
accessible system sizes. The NIM on the other hand shows a clear
(exponential at high temperature) decay of $\bar{D}_e$ with system
size indicating diffusion. This behavior persists to temperatures
higher than the typical energy scales in the problem and {\em in
fact even up to infinite temperature}. These results are
interesting since they show that the system is aware of its
integrability or non-integrability even up to infinite
temperature. $D_e$ vanishes rapidly in the thermodynamic limit in
both systems (even without the flux twist). This shows in
particular that the IM despite having an infinite conductivity at
$\omega=0$ is not a superconductor. It is a curious fact that
$D_e$ is positive at small system sizes in the IM and negative in
the NIM. The situation is reversed without a flux twist. The
transport stiffnesses in small systems can in general have either
signs depending on the microscopics~\cite{stafford}.

We now outline a statistical study of $\bar{D}_e$ in the two
systems. Since the Meissner stiffness $D_e$ is zero at large $L$,
$\bar{D}_e$ is given by Eqn.~\ref{stiffshort}. To enable a
comparison of $\bar{D}_e$ in the IM and the NIM at a given $L$, we
remove the dependence on microscopic parameters (i.e. normalize)
by dividing by $\Gamma_e$. We also normalize $\bar{D}_e^d$ and
$\bar{D}_e^{nd}$ in the same way. For the largest system sizes
($L=18$), we find that the normalized $\bar{D}_e$ is about 8-15
times larger in the IM than the NIM for the values of $\beta$
considered. One can now ask the question we set out to answer in a
slightly different from: Are the large current carrying states
which give a large (normalized) $\bar{D}_e$ in the IM compared to
the NIM in the degenerate states in the former, or in the
non-degenerate ones? This can be answered by comparing the
magnitudes of the normalized $\bar{D}_e^d$ and $\bar{D}_e^{nd}$.

Motivated by this question, we have studied the statistics of the
number of degeneracies and the the current matrix elements in the
two systems. We find that at $L=18$ in the IM, about 10$\%$ of the
states are pairwise degenerate with a very negligible fraction of
higher order degeneracies. Further, most of the degeneracies occur
among states with total lattice momentum 0 or $\pi$. The ratio of
the nrmalized $\bar{D}_{e}^{d}/\bar{D}_e^{nd}$ is at most 0.2 for
the values of $\beta$ investigated. On the other hand, the ratio
of $\bar{D}_{e}^{nd}$ between the IM and the NIM is about 8-15.
Thus, we come to the following interesting conclusion: Even though
the IM has degeneracies, their contribution to the charge
stiffness is not significant. {\em The large current carrying
states are primarily among the non-degenerate states.} We also
note that the eigenvalues of the current operator in a given
degenerate sector in the IM, are always roughly of the same
magnitude. We reiterate that the term non-degenerate is used here
only for states with no degeneracies {\em within} a sector of
total momentum $q$. Thus, it appears the presence of dynamical
symmetries does not have a direct effect on the charge stiffness
through the creation of degenerate states. A more detailed account
of the statistics alluded to above will be presented elsewhere.

We now focus on transport at finite frequency ($\omega$) and $q$.
The $q$ and $\omega$ dependent conductivities are given by
\begin{equation}
A_\alpha(q, \omega) = c(\omega) \sum_{p, \epsilon_n \neq \epsilon_m} p_n
|\langle n|J_\alpha(q)|m\rangle |^2
\delta(\epsilon_m -\epsilon_n - \hbar \omega), \label{Kubo}
\end{equation}
where $ c(\omega)= \frac{\pi}{L}\left(\frac{1-e^{-\beta
\omega}}{\omega}\right)$,  $A_e(q, \omega )=\sigma(q,\omega)$ and
$A_E(q, \omega)=\kappa(q,\omega)$. These conductivities can also
be related to density-density correlation functions of the charge
and energy~\cite{zotos2}. We choose a small value of $\beta=0.001$
since for a calculation of this sort, numerical exact
diagonalization is most efficient only at very high
temperatures~\cite{mukerjee}. The irrational flux twist turns out
to be unimportant to the results here. $\sigma(q,\omega)$ and
$\kappa(q,\omega)$ as functions of $\omega$ (in units of the
hopping $t$) for different values of $q$ (going from 0 to 8) for
an $L=16$ IM are shown in Fig.~\ref{diffmom}. $\kappa(q=0,\omega)
\propto \delta(\omega)$ since $[J_E(q=0),H]=0$ but
$[J_e(q=0),H]\neq 0$ and thus $\sigma(q=0,\omega)$ has some
structure at $\omega \neq 0$. A more interesting feature is that
$\kappa(q,\omega)$ is non-zero only within a band of frequencies
for small values of $q$ and goes to zero abruptly at the
boundaries of the band. Moreover, this band shifts to higher
frequencies with increasing $q$. The dispersion of these banded
modes appears to be roughly linear (at small $q$), consistent with
considerations of ballistic transport. The situation is similar to
the case of free fermions where ballistic transport causes a
similar banded structure in $\kappa(q,\omega)$. The common feature
of this IM and free fermions is integrability, which it appears is
strongly associated with the concept of ballistic transport.
$\sigma(q,\omega)$ does not display the same banded feature as
$\kappa(q,\omega)$ as prominently, presumably due to the fact that
$[J_e(q=0),H]\neq 0$, where the analogy with free fermions does
not apply. Fig.~\ref{colcont} shows contour plots of
$\sigma(q,\omega)$ and $\kappa(q,\omega)$ to better illustrate the
banded nature of $\kappa(q,\omega)$ and enable comparison to
$\sigma(q,\omega)$.

For the sake of comparison we also present numerical data for
$\sigma(q,\omega)$ and $\kappa(q,\omega)$ of the NIM. The plots of
these quantities are shown in Fig.~\ref{diffmomn}. $[J_E(q=0),H]
\neq 0$ here and $\kappa(q=0,\omega)$ is non-zero at $\omega \neq
0$. Both $\sigma(q=0,\omega)$ and $\kappa(q=0,\omega)$ display a
finite non-analytic singularity at $\omega =0$, which has been
attributed to diffusion and non-linear hydrodynamics for the
former~\cite{mukerjee}. In this system $\kappa(q,\omega)$ does not
display the banded behavior of the IM and gradually goes to zero
with increasing $\omega$ for all $q$. This is indicative of
diffusion in this system, which we have verified by also computing
the density-density correlators directly. The features (bumps) at
large values of $q(=5-8)$ and $\omega$ also appear in these
correlators indicating the presence of interesting overdamped
excitations, which will be investigated in detail elsewhere.

To conclude, we have demonstrated that the IM shows several signs
of ballistic behavior as opposed to diffusion in the NIM. This has
been illustrated through an extensive numerical calculation of
finite temperature transport stiffnesses and dynamical
conductivities. Futher, the contribution of degenerate states
coupled by the current has been shown to be insignificant to the
Drude weight in the IM. The NIM seems to possess overdamped
excitations at large momenta

The authors thank J. O. Haerter, D. A. Huse, J. E. Moore, M. R.
Peterson and V. Oganeysan for discussions. S. M. thanks the DOE
for support and the IBM SUR program. B.S.S. was supported by the
DOE, BES under grant DE-FG02-06ER46319
\bibliographystyle{unsrt}
\bibliography{bigbib}
\end{document}